\documentclass[11pt,aps]{article}
\usepackage[sfdefault]{AlegreyaSans} 
\usepackage{geometry}
\usepackage{setspace}
\usepackage{titlesec}
\PassOptionsToPackage{hyphens}{url}\usepackage{hyperref}
\usepackage{graphicx}
\usepackage{amsmath}
\usepackage{bm}
\usepackage{wrapfig}
\usepackage{xcolor}
\usepackage{amssymb}
\usepackage{authblk}
\usepackage[export]{adjustbox}
\usepackage{subcaption}
\usepackage{tikz} 
\usetikzlibrary{matrix}
\usepackage[numbers,sort&compress]{natbib}

\usepackage{comment}
\usepackage{upgreek}

\titleformat{\section}{\normalfont\fontsize{14}{16}\bfseries\sffamily}{\thesection}{1em}{}
\titleformat{\subsection}{\normalfont\fontsize{12}{14}\bfseries\sffamily}{\thesubsection}{1em}{}
\titleformat{\title}{\normalfont\fontsize{12}{14}\bfseries\sffamily}{}{0em}{}

\title{Predicting Brain Morphogenesis via Physics-Transfer Learning}

\author[1]{Yingjie Zhao}
\author[2]{Yicheng Song}
\author[2]{Fan Xu$^\ast$}
\author[1]{Zhiping Xu$^\ast$}
\affil[1]{Applied Mechanics Laboratory, Department of Engineering Mechanics, Tsinghua University, Beijing 100084, China}
\affil[2]{Institute of Mechanics and Computational Engineering, Department of Aeronautics and Astronautics \& College of Intelligent Robotics and Advanced Manufacturing, Fudan University, Shanghai 200433, China}
\affil[$^\ast$]{Corresponding authors: Fan Xu (fanxu@fudan.edu.cn), Zhiping Xu (xuzp@tsinghua.edu.cn).}

\date{}

\newlength{\tempdima}
\newcommand{\rowname}[1]
{\rotatebox{90}{\makebox[\tempdima][c]{\textbf{#1}}}}

\newcommand\zfig[1]{{\color{violet}#1}}

\usepackage[T1]{fontenc}
\usepackage{array}
\usepackage{booktabs}
\usepackage{float}

\begin{document}

\maketitle

\begin{abstract}
Brain morphology is shaped by genetic and mechanical factors and is linked to biological development and diseases.
Its fractal-like features, regional anisotropy, and complex curvature distributions hinder quantitative insights in medical inspections.
Recognizing that the underlying elastic instability and bifurcation share the same physics as simple geometries such as spheres and ellipses, we developed a physics-transfer learning framework to address the geometrical complexity.
To overcome the challenge of data scarcity, we constructed a digital library of high-fidelity continuum mechanics modeling that both describes and predicts the developmental processes of brain growth and disease.
The physics of nonlinear elasticity from simple geometries is embedded into a neural network and applied to brain models. 
This physics-transfer approach demonstrates remarkable performance in feature characterization and morphogenesis prediction, highlighting the pivotal role of localized deformation in dominating over the background geometry.
The data-driven framework also provides a library of reduced-dimensional evolutionary representations that capture the essential physics of the highly folded cerebral cortex.
Validation through medical images and domain expertise underscores the deployment of digital-twin technology in comprehending the morphological complexity of the brain.
\end{abstract}

\clearpage
\newpage


Brain development is a multi-scale physical process that spans from molecular gene expression and protein folding to cellular division, differentiation, and migration, culminating in macroscopic cortical folding (\zfig{Fig. S1})~\cite{silbereis2016Neuron,llinares2019NRN}. 
Genetic programs that direct anisotropic growth interact with mechanical forces to sculpt the brain’s structure. These processes unfold over time and space, giving rise to the evolving morphology observed throughout life~\cite{silbereis2016Neuron,llinares2019NRN}.

The brain’s geometry is highly intricate.
Neurons are embedded in a dense, interconnected network, and its surface is patterned with gyri and sulci~\cite{park2013Science}, which are key features intimately linked to cognition and function~\cite{kriegeskorte2021NRN,pang2023Nature}.
Morphological patterns also serve as sensitive indicators of neurodevelopment, disease progression, and aging~\cite{oegema2020NRN,hutton2009Neuroimage}.
Understanding how these features emerge and evolve could illuminate mechanisms of morphogenesis, functional specialization, and pathology.
Progress hinges on integrating high-resolution observations with models capable of predicting individualized developmental trajectories.

Magnetic resonance imaging (MRI) is the primary tool for capturing in vivo brain morphology~\cite{makropoulos2018Neuroimage}.
Yet, MRI studies are predominantly cross-sectional, with longitudinal datasets remaining scarce (\zfig{Fig. 1})~\cite{bethlehem2022Nature}.
Resolution limits hinder the capture of fine cortical folding patterns, which are non-uniform, hierarchical, and individual-specific.
Moreover, MRI provides geometry but not underlying physiological drivers, such as growth stresses responsible for folding (\zfig{Fig. 1})~\cite{xu2009BMM,xu2010JBE}, making predictive diagnosis and prognosis difficult~\cite{arbabshirani2017NI,varoquaux2022NPJDM}.

Population-averaged models offer valuable baselines~\cite{severino2020Brain} but obscure the marked variability in neuroanatomical maturation between individuals.
This variability limits the diagnostic value of normative curves and hampers early detection of atypical development~\cite{alenya2022BrainMulti,lorenzo2025PR}.
Computational approaches capable of forecasting individual-specific morphological trajectories could enable earlier intervention, better disease subtype stratification, and personalized therapeutic design.

Continuum mechanics offers a natural multiscale framework for brain morphogenesis~\cite{darayi2021review}, incorporating genetic and biochemical cues via growth tensors, and mechanical properties, such as stiffness and viscoelasticity, via constitutive laws.
These elements together drive local expansion, deformation, and mechanical instabilities.
High-fidelity simulations can recover characteristic patterns such as sulci, gyri, and bifurcations~\cite{tallinen2014PNAS,tallinen2016NaturePhysics,budday2018IJSS,da2021NeuroImage,wang2021SR}.
However, the brain’s geometric complexity makes such simulations computationally expensive, and the paucity of high-quality MRI data further impedes comprehensive digital reconstructions.

Generative artificial intelligence (AI) offers an appealing route to augment datasets, but purely data-driven approaches trained only on morphological data struggle with generalization and physical plausibility~\cite{puglisi2025PR}.
Embedding physical priors into AI can alleviate these issues~\cite{karniadakis2021NRP}, whether by enforcing governing equations directly in the loss function~\cite{raissi2019JCP} or by incorporating symmetry and conservation constraints~\cite{batzner2022NC}.
For brain morphogenesis, however, direct imposition of strong physical constraints is computationally costly~\cite{mcgreivy2024NMI}, and simple symmetries are absent.
The challenge lies in capturing growth-driven elastic instabilities and nonlinear material behaviors within feasible AI architectures.

Here, we introduce a physics-transfer (PT) learning framework~\cite{zhao2024PR} for modeling brain morphological development.
The key idea is to first learn the nonlinear deformation and bifurcation physics from simplified geometries using high-fidelity continuum mechanics, then transfer this knowledge, encoded in neural network (NN) weights, to predict individual brain evolution.
We construct a dense, high-coverage digital library of spherical and ellipsoidal growth models to capture universal bifurcation behaviors, and then apply the learned physics to human brain geometries.
Analysis of this data-driven model uncovers the physical principles encoded within the NN, emphasizing the predominant influence of cortical folding over background configurational curvature.
These results underscore the pivotal role of localized deformation in driving morphogenesis.
Predictions are validated against MRI-derived data, achieving high accuracy while offering interpretable insights.

Our results demonstrate that PT learning bridges data scarcity and physical complexity, enabling realistic, temporally continuous reconstructions of brain development.
This approach paves the way for digital twin models of the brain, linking morphology, mechanics, and function across scales.

\section*{Results}

\subsection*{Brain models}

Brain morphological development arises from the interplay of physical complexity, encompassing genetic regulation and mechanical forces, and geometric complexity.
The multiscale physics governing this process can be described within the framework of continuum mechanics, in which the growth tensor serves as a key physiological parameter for quantifying tissue growth kinetics (see \zfig{Methods} for details)~\cite{tallinen2016NaturePhysics,striedter2015review,darayi2021review,budday2018IJSS,da2021NeuroImage,alenya2022BrainMulti}.
These parameters bridge microscale cellular behaviors and microstructural organization, enabling multiscale models that capture morphological instabilities\cite{garcia2021NC}.
Within this framework, the nonlinear deformation physics underlying cortical pattern formation, manifested in sulci, gyri, and bifurcation structures, can be faithfully reproduced.
When coupled with machine learning (ML), these mechanical models offer a foundation for digital brain twins to inform patient-specific clinical decisions and precision medicine, though further refinement is required~\cite{corral2020EHJ,alenya2022BrainMulti}.

Yet, modeling brain morphological development remains challenging, owing to the intrinsic complexity of neurodevelopment and the scarcity of longitudinal MRI datasets.
Finite element analysis (FEA) is widely used to simulate brain growth and generate synthetic data, but geometrical nonlinearity leads to low computational efficiency and poor convergence (\zfig{Fig. 2a})~\cite{tallinen2016NaturePhysics}.
Consequently, most studies avoid direct simulation of realistic brain geometries, focusing instead on idealized two-dimensional (2D) shells or three-dimensional (3D) spheres and ellipsoids~\cite{darayi2021review,budday2018IJSS,da2021NeuroImage,wang2021SR}.
This geometric complexity hampers the construction of a comprehensive digital brain library from limited MRI datasets, resulting in severe data sparsity (\zfig{Fig. 2b}).

Interestingly, even such simplified geometries exhibit spatiotemporal features reminiscent of brain morphology, including vertex-ridge networks and bifurcation patterns.
Motivated by these parallels, we construct a high-density, high-coverage digital library of morphological evolution using simplified geometries, and transfer the learned physics to complex brain structures.
This strategy overcomes data sparsity and enables physics-informed modeling of realistic brain development (\zfig{Fig. 2c}).

\subsection*{Digital libraries}

Brain development data are both distinct and scarce. Even the most comprehensive MRI repository, comprising $123,984$ scans, contains only $\sim 10 \%$ longitudinal records, often acquired at irregular intervals~\cite{bethlehem2022Nature}.
For brain morphology specifically, a single atlas is currently available~\cite{ciceri2024NI}.
This scarcity of high-quality, temporally consistent datasets poses a critical barrier to modeling and understanding brain morphogenesis.

To address this gap, we construct digital libraries of morphological patterns based on spheres and ellipsoids (\zfig{Fig. S2}, and see \zfig{Methods} for details).
These are modeled using a representative core-shell framework~\cite{tallinen2014PNAS,wang2021SR,xu2022NCS,yin2008PNAS,lin2020EML,wang2011SoftMatter}, widely employed to investigate mechanical instabilities in cortical folding~\cite{tallinen2016NaturePhysics,striedter2015review,darayi2021review,budday2018IJSS,da2021NeuroImage,alenya2022BrainMulti}.
In this framework, the outer shell represents the cerebral cortex (gray matter) and the inner core the white matter.
Both are modeled as modestly compressible, hyperelastic Neo-Hookean materials with distinct growth rates~\cite{tallinen2016NaturePhysics}.
Following experimental observations~\cite{fischl2000PNAS,chang2007abnormal,xu2010JBE,dervaux2008PRL,budday2015JMBBM}, cortical thickness varies from $0.03-1.63\ \mathrm{mm}$, encompassing both normal and abnormal human measurements, while the relative shear modulus ($G_{\mathrm{grey}}/G_{\mathrm{white}}$) spans $0.65-1$.

In neurodevelopment, biomarkers are quantifiable anatomical or physiological features indicative of underlying processes or deviations.
Cortical thickness is a key biomarker, influencing brain morphology and serving in the diagnosis of cortical malformations, including lissencephaly, polymicrogyria, and focal cortical dysplasia~\cite{wang2021SR}.
To simulate the growth stresses driving pattern evolution, we adopt the tangential growth (TG) model, which captures the cellular mechanisms that generate morphological instabilities~\cite{tallinen2014PNAS,tallinen2016NaturePhysics,llinares2019NRN}.

For predictive modeling, ML models are trained on datasets derived from these simple geometries, then applied zero-shot to brain morphology, avoiding the high computational costs of direct modeling, which must resolve both global geometry (structure, shape, size) and local features (cortex, subcortical structures, ventricles) (see \zfig{Methods} for details).
Sampling from the morphological development space of simple geometries is markedly more efficient than from brain tissues (\zfig{Fig. 2c}), while preserving spatiotemporal similarities to brain development.
This enables robust generalization to out-of-distribution (OOD) brain data.

The PT approach proceeds by training on in-distribution (ID) simple geometries with an enriched sampling space, allowing the model to learn essential physical principles underlying morphological change.
These learned representations are then transferred to brain models via NNs with fixed weights that encode the core physics.
This strategy ensures that the robustly generalizes from the simple geometries (ID) to that of the more complex and data-scarce brain morphologies (OOD).
Verified predictive accuracy and validation with experimental data demonstrate the success in resolving the accuracy-performance dilemma, imposed by geometric complexity.

\subsection*{Predictions of curvature maps and 3D morphology}

We begin with a simple scenario where the goal is to predict biomarkers such as the curvature map from 3D morphological data, formulated as a regression task (\zfig{Figs. 3a}, \zfig{S3}, and see \zfig{Methods} for details).
Using an encoder-decoder graph neural network (GNN) architecture (\zfig{Fig. S4}), we represent morphology as a graph where each node encodes features including spatial coordinates and normal vectors.
The model predicts local curvature values, which have been demonstrated as key descriptors of morphological complexity~\cite{zhao2022Patterns,zhao2024CRPS}, closely associated with neurodevelopmental trajectories, and serve as indicators of neurological function and disease.

As a control, we trained a statistical learning (SL) model that excluded normal vectors from the inputs, retaining only morphological data and thereby limiting its ability to capture the physics underlying curvature, which depends on nonlinear elasticity (see \zfig{Methods} for details).
Both SL and PT methods accurately predict curvature, computed as the sum of absolute mean and Gaussian curvatures, $\lvert H \rvert + \lvert K \rvert$ (\zfig{Fig. 3b}).
However, when applied to ellipsoids and experimental human brain data obtained from experimental MRI images, PT learning significantly outperforms SL, achieving mean absolute errors (MAEs) between the predicted and ground-truth curvature maps (reconstructed from MRI images) below $0.5$ mm$^{-1}$, compared to $7.01$ mm$^{-1}$ for the latter (\zfig{Figs. 3c} and \zfig{3d}).
This validation using model ellipsoidal data and experimental brain data underscores the outstanding generalizability provided by the physical principles embedded within the NNs.

We then extend to a spatiotemporal prediction problem, to forecast future morphological development from existing data.
This capability is vital for early diagnosis, disease monitoring, and personalized treatment planning.
Due to the scarcity of longitudinal brain data, models are pretrained on simple geometry libraries before transferring learned physics to brain morphologies.

An encoder-decoder GNN~\cite{pfaff2020ICLR} processes morphology graphs whose nodes carry spatial and normal features, while edges represent mesh connectivity or contact events, characterized by nodal distances in reference and current configurations.
The encoder module embeds input features into a latent space through message-passing in graph convolution.
The decoder module decodes latent features into nodal acceleration at each node.
An integration module based on Newton's second law is embedded into the network to compute the position increments of the nodes and update their positions at the next time step~\cite{pfaff2020ICLR}.
During inference, an autoencoder or iterative rollout is employed, enabling a digital twin approach to predict brain morphological development for medical applications (\zfig{Fig. 3e}).

For one-step prediction (i.e., predicting brain morphology one week ahead), the PT models achieve high accuracy, with MAE between prediction and the ground truth reconstructed from MRI images below $0.01$ (\zfig{Figs. 3f} and \zfig{S5}).
Although errors increase over longer horizons (e.g., five weeks), visual inspection confirms preservation of key structural features.
Clinical relevance dictates error thresholds.
Global metrics like head circumference allow higher tolerance, whereas diagnoses sensitive to local detail, such as schizophrenia, require stricter limits~\cite{budday2014SR}.
Beyond these thresholds, manual corrections guided by the digital twin framework become necessary (\zfig{Figs. 3e, f}).

Finally, we quantify the relationship between latent space features and model generalizability to enable a priori estimation of model performance (\zfig{Fig. S6a}, and see \zfig{Methods} for details).
The latent-space distance between training spheres and testing brain data positively correlates with prediction error (\zfig{Figs. S5} and \zfig{S6b}).
Short-term inputs (1-step) exhibit minimal latent distance and lowest errors, whereas longer-term inputs (7-step) show increased distance and degraded generalization (\zfig{Figs. S6c} and \zfig{S6d}).
This latent distance metric provides an interpretable estimate of prediction uncertainty and serves as an early warning for potential model failure.

\section*{Discussion}
\subsection*{Alignment with medical images}

MRI brain imaging provides crucial, non-invasive insights into neurological anatomy and function, aiding diagnosis~\cite{severino2020Brain,ciceri2024NI}.
However, predictive modeling of neurological diseases is limited by the scarcity and low fidelity of MRI data~\cite{arbabshirani2017NI,varoquaux2022NPJDM}.
In particular, longitudinal MRI scans for individuals are sparse, hindering accurate tracking of brain morphological development. 
Moreover, MRI primarily captures geometric features and lacks detailed information on biomechanical forces driving cortical folding, constraining morphology prediction from images alone (\zfig{Fig. 1}).

By integrating these limited clinical images with a richly sampled digital library and PT, we overcome these challenges.
Our digital library provides dense, comprehensive data that encode nonlinear elasticity and folding mechanics within ML models.
This fusion enables PT models to accurately predict brain morphology using sparse imaging data.
Furthermore, augmenting these data with multimodal inputs, such as genetic, cellular, and biomechanical information, promises to enhance predictive diagnostics and data integration in neurodevelopment.

\subsection*{Reduced-dimensional representations}

Reduced-dimensional models are essential for advancing scientific understanding and engineering of complex systems.
They distill underlying physical principles from observational data, deepening insight and enabling tasks such as control, design, and decision-making.
Compared to full-order models, reduced models offer dramatic gains in computational efficiency, often reducing dimensionality by several orders of magnitude, thereby supporting scalability and real-time simulation.
Yet, constructing such models remains challenging, especially for systems governed by nonlinear, multiscale, and non-equilibrium physics.

Brain morphological development exhibits evolutionary patterns of vertex–ridge networks resembling those seen in spherical growth and shrinkage (\zfig{Fig. 4a}), suggesting that a reduced-dimensional description of these dynamics is feasible.
Quantitative metrics, including vertex and ridge counts, track morphological evolution over time (\zfig{Figs. 4b, c, and Supplementary Note S1})~\cite{ester1996KDD}.
However, deriving governing equations is complicated by nucleation, annihilation, mutual interactions, and coupling with the surrounding tissue environment.
Our PT framework produces rich, high-fidelity reduced-dimensional representations of brain morphology, laying the groundwork for future data-driven discovery of dynamical equations governing vertex–ridge network evolution~\cite{chen2021NC}.

\subsection*{Model interpretability}

On the other hand, AI and ML provide powerful tools for reduced-dimensional modeling by enabling automatic extraction of governing dynamics from observational data.
From a connectionist viewpoint, NNs approximate complex mappings between system variables, capturing intricate, high-dimensional behaviors~\cite{hinton1990connectionist}.
Despite the expressive power of well-trained neural networks, their internal representations often lack interpretability, making it challenging to extract physically meaningful laws and underscoring the persistent gap between connectionist learning and the understanding required for scientific discovery.

To close this gap, we apply dimensionality reduction to the internal representations of well-trained NNs to uncover interpretable physical insights emerging organically within the network (\zfig{Figs. 4d, e, and Supplementary Note S1})~\cite{la2024PR}.
For brain morphology, the PT-trained model exhibits similar weight distributions ($p(\theta|\mathcal{D}_{\rm L}^{'}) \approx p(\theta|\mathcal{D}_{\rm H}^{'})$), across datasets of varying geometrical complexity (spheres and ellipsoids), evaluated by mean layer-wise weights evaluated through the mean value $\mu$ (\zfig{Fig. 4d}).
In contrast, SL models show marked differences in parameter distributions between low- and high-complexity datasets ($p(\theta|\mathcal{D_{\rm L}}) \neq p(\theta|\mathcal{D_{\rm H}})$) as compared to PT learning (\zfig{Fig. 4e}).
This consistency in PT weight distributions reflects improved generalizability by aligning physics across domains~\cite{ben2010DA}.

Neuron activation patterns further illuminate PT’s superior OOD performance (\zfig{Figs. 4f} and \zfig{4g}), consistent with curvature mapping results (\zfig{Fig. 3d}).
PT models trained on spherical data produce activation states resembling those elicited by brain morphology data (\zfig{Fig. 4f}), whereas SL models exhibit divergent activations across these domains (\zfig{Fig. 4g}).

Brain morphogenesis features structural hierarchies spanning multiple spatial scales, including large-scale configuration curvature and smaller-scale deformation curvature such as wrinkling wavelength and radius (\zfig{Fig. S7a}).
To quantify the relative importance of these scales, we leverage ML models coupled with information bottleneck (IB) theory~\cite{tishby2000IB,tishby2015IB}.
The latent features in ML models encode curvatures at different scales in distinct ways, enabling the model to disentangle and assess their respective contributions to morphogenesis.
Within this framework, the mutual information between input and latent representations, denoted as $I(X;H)$, quantifies information compression during training (\zfig{Fig. S7b}).
As training progresses, $I(X;H)$ converges, indicating that the model has distilled essential latent features that capture the fundamental physical principles of morphogenesis (\zfig{Figs. S7c, d, and Supplementary Note S2}).
Monitoring information compression shows that background configurational curvature undergoes greater compression than deformation curvature, indicating that deformation curvature plays a more critical role in morphogenesis (\zfig{Fig. S7d and Supplementary Note S2}).

\subsection*{Medical applications}

Clinical experts diagnose brain malformations and diseases by integrating MRI data with clinical symptoms and patient history~\cite{oegema2020NRN}.
Structural MRI and accompanying metadata enable identification of a spectrum of conditions.
White matter abnormalities underpin diagnoses of schizencephaly, gray matter heterotopia, and focal cortical dysplasia, cortical morphology reveals malformations such as polymicrogyria, cobblestone malformation, and lissencephaly, and overall brain volume anomalies diagnose megalencephaly and microcephaly~\cite{severino2020Brain}. 
Aggregated across the lifespan, MRI data inform reference brain charts that delineate normative and pathological developmental trajectories via neuroimaging biomarkers derived from structural and morphological metrics~\cite{bethlehem2022Nature}.
These biomarkers are critical for diagnosing neurodevelopmental and neurodegenerative disorders, including Alzheimer’s disease, attention-deficit/hyperactivity disorder (ADHD), autism spectrum disorder, anxiety disorders, bipolar disorder, major depressive disorder, and schizophrenia~\cite{bethlehem2022Nature}. 
Generative AI models further hold promise for modeling longitudinal MRI changes and disease progression, advancing data-driven, personalized clinical decision-making~\cite{puglisi2025PR}.

Conventional approaches focus largely on anatomical geometry extracted from MRI data.
Our PT framework, by contrast, integrates the physics of nonlinear elasticity governing brain morphogenesis into predictive models supported by a curated digital library.
The resulting curvature maps, robust morphological descriptors, serve as biomarkers of developmental brain disorders, reflecting the established links between aberrant curvature and pathology~\cite{budday2014SR}.
Within a digital twin paradigm, patient-specific brain morphology reconstructed from MRI can be input into the PT model to assess current structural states and forecast future development (\zfig{Fig. 5a}).
Detection of a trajectory toward pathological morphological bifurcations may prompt timely clinical intervention, with ongoing MRI data facilitating iterative model refinement for real-time monitoring and prognostic evaluation.

Unlike standardized data domains such as computer vision (CV) or natural language processing (NLP), brain MRI data exhibit significant heterogeneity due to varying imaging protocols, scanner types, and population diversity.
Moreover, privacy and ethical constraints limit data sharing and integration, posing unique challenges.
Progress in AI-driven modeling of brain development and predictive diagnostics thus depends on establishing open, well-curated, interoperable databases coupled with comprehensive digital libraries dedicated to brain morphology (\zfig{Fig. 5b}).

\subsection*{Limitations}

Our model exhibits limitations in predicting brain morphological changes arising from rare events such as severe traumatic brain injuries (e.g., motor vehicle accidents, gunshot wounds, high-impact falls, explosions), uncommon infectious diseases (e.g., Creutzfeldt-Jakob disease, cerebral malaria), and exposure to extreme environments such as spaceflight (\zfig{Fig. 5c})~\cite{hellstrom2017BI,collie2001CR,rezaei2024BIB}.
Accurate diagnosis of brain conditions linked to these infrequent occurrences requires further evaluation that integrates patient history and clinical presentation under expert medical guidance.

The deformation physics in our digital library is based on established constitutive models of brain development, which involve complex formulations and parameter uncertainties~\cite{darayi2021review}.
Improved understanding through theoretical, experimental, and ML-driven models will enhance accuracy~\cite{linka2023AB}.
By integrating diverse state-of-the-art models, the PT framework surpasses reliance on any single FEA approach.
This also applies to growth factors linked to biomolecular kinetics~\cite{darayi2021review,llinares2019NRN}, where multi-scale data integration can deepen insight into brain morphogenesis.

The predictive accuracy of PT depends on the quality of its digital libraries, constrained by existing theoretical and experimental knowledge.
Expanding MRI datasets, such as via the Developing Human Connectome Project~\cite{makropoulos2018NI}, could further improve predictions, beyond this study’s scope.

\section*{Conclusion}

We present a PT learning framework that integrates high-fidelity continuum mechanics modeling with NNs to advance the understanding and prediction of brain morphogenesis.
Exploiting shared biophysical principles across simple and complex geometries, the approach overcomes challenges of data scarcity and geometric complexity.
Unlike generative models that map distributions statistically, PT learning synthesizes developmental trajectories by embedding the physics of spatiotemporal evolution derived from spherical and ellipsoidal geometries.
The resulting low-dimensional representations isolate the essential mechanisms of cortical folding in brain development, with large-scale configurational curvature shown to be irrelevant.
Validation against medical imaging demonstrates the framework’s predictive fidelity.
These results underscore the potential of physics-informed digital twins to bridge mechanistic modeling and data-driven inference, enabling deeper insight and improved forecasting of brain developmental dynamics.

\clearpage
\newpage

\section*{Methods}

\subsection*{Continuum modeling}

Brain development is regulated by genetic, molecular, cellular, and mechanical factors across multiple spatiotemporal scales~\cite{klingler2021Science,llinares2019NRN}, and the differential tangential growth hypothesis is commonly used~\cite{tallinen2016NaturePhysics,klingler2021Science,llinares2019NRN}.
Finite element analysis (FEA) can model morphological evolution during brain growth at the continuum level~\cite{tallinen2016NaturePhysics,tallinen2014PNAS,darayi2021review,budday2018IJSS,wang2021SR}.
Tangential growth (TG) of the outer gray matter is faster than the inner white matter, and its implementation in FEA is known as the TG model~\cite{tallinen2016NaturePhysics}.
Compression resulting from the mismatch in deformation may then lead to mechanical instabilities of the brain surface, forming characteristic sulci and gyri structures~\cite{tallinen2014PNAS,tallinen2016NaturePhysics,striedter2015review,darayi2021review,wang2021SR,budday2018IJSS,da2021NeuroImage}.

In continuum modeling, the reference configuration can be mapped to the current one through the deformation gradient tensor as
\begin{equation}\label{eq4}
	\textbf{F} = \textbf{F}^{\rm e}\cdot\textbf{G},
\end{equation}
where $\textbf{F}^{\rm e}$ is the elastic deformation gradient and $\textbf{G}$ is the growth term.
In the TG model, the growth tensor $\textbf{G}$ is
\begin{equation}\label{eq5}
	\textbf{G} = g\textbf{I} + (1-g)\hat{\textbf{n}}\otimes\hat{\textbf{n}},
\end{equation}
where $\hat{\textbf{n}}$ is the surface normal of the reference configuration, $\textbf{I}$ is the unit tensor, and
\begin{equation}\label{eq6}
	g = 1 + \frac{\alpha_{t}}{1+e^{10(\frac{y}{T}-1)}}
\end{equation}
is the growth coefficient, where $\alpha_{t}$ controls the magnitude of local cortical expansion.
The constant factor of $10$ in the exponential term governs the steepness of the sigmoidal transition in $g$, producing a sharp yet smooth change near $y = T$ that delineates the gray matter layer from the underlying white matter~\cite{tallinen2014PNAS,tallinen2016NaturePhysics,wang2021SR}.
There is a smooth transition from the surface of the gray matter layer to the white matter layer with a gradually decreasing growth coefficient.
$y$ is the distance to the surface, and $T$ is the thickness of the cortex.
The brain is modeled as a nonlinear neo-Hookean hyperelastic material, where the strain energy density is
\begin{equation}\label{eq7}
	W = \frac{G}{2}[\rm{Tr}(\textbf{F}^{\rm e}\textbf{F}^{\rm eT})\emph{J}^{-2/3}-3] + \frac{\emph{K}}{2}(\emph{J}-1)^2,
\end{equation}
where $G$ is the shear modulus, $J$ is the determinant of Jacobian matrix, $K$ is the bulk modulus.

For brain growth, a core-shell structure with a spherical geometry is used for its simplicity.
The outer radius is $10$ mm and the shell thickness ranges from $0.03$ to $1.63$ mm, which are determined from the measurements of abnormal and normal human cerebral cortices~\cite{fischl2000PNAS,wang2021SR}.
$4$-node tetrahedral elements with a density of $10^6$ tetrahedra/cm$^3$ for discretization with the convergence confirmed~\cite{tallinen2016NaturePhysics,wang2021SR}.
The morphogenesis of brains is triggered by internal elastic stresses generated from differential core-shell growth.
The interaction between surfaces is modeled with an energy penalty via vertex-triangle contact, which prevents the nodes from penetrating the faces of elements~\cite{ericson2004contact}.
An explicit solver is used to minimize the total (elastic and contact) energy of the quasi-static system.
The time step $\Delta t = 0.05a\sqrt{\rho/\emph{K}}$ is set to ensure the convergence, where $a$ is the mesh size and $\rho$ is the mass density~\cite{belytschko2014FEA}.

Assigning material models and parameters to brain tissue regions is challenging due to intra-regional variability and differences across individual brains.
Additionally, properties change with development or aging.
The alternative approach taken here, which is the current state of the art, involves assigning `typical' properties for a tissue type and age, using experimental data that closely approximate the specific loading conditions.
The bulk modulus of the core and shell is $5$ times the shear modulus~\cite{tallinen2016NaturePhysics}.
Following the experimental evidence, the relative shear modulus ($G_{\rm shell}/G_{\rm core}$) ranges from $0.65$ to $1$~\cite{budday2015JMBBM}.

\subsection*{Machine learning models}

Models with varying complexities ($\mathcal{C}$) in multi-scale modeling show different parameter distributions, denoted as $p(\theta|\mathcal{C})$, where $\theta$ are the model parameters and $p(\cdot|\cdot)$ represents the conditional probability.
The parameter distributions $p(\theta|\mathcal{D})$ in a specific ML context typically depend on data complexities ($\mathcal{D}$).
Generally, data with low ($\mathcal{D}_{\rm L}$) and high ($\mathcal{D}_{\rm H}$) complexities can exhibit different distributions, that is

\begin{equation}\label{eq1}
p(\theta|\mathcal{D_{\rm L}}) \neq p(\theta|\mathcal{D_{\rm H}}),
\end{equation}

which limits the transferability and extrapolation of models trained on data with different complexities.

In certain situations, the physics ($\mathcal{P}$) underlying the dataset ($\mathcal{D}$) can facilitate extrapolation through a physics-transfer (PT) approach.
If such a physical relationship exists between the features ($\mathbf{x}$) and the target ($\mathcal{O}$) in $\mathcal{D}$, then we have

\begin{equation}\label{eq1}
\mathbf{x} \xrightarrow{\mathcal{P}} \mathcal{O},
\end{equation}
\begin{equation}\label{eq1}
\mathbf{x} \cap \mathcal{O} = \mathcal{D}^{'} \subset \mathcal{D},
\end{equation}

where $\mathcal{D}^{'}$ represents a space of reduced dimensions.
Specific ML models ($h \in \mathcal{H}$) can be designed to learn the underlying physics of the data.
Models trained on data of varying fidelities tend to share a similar parameter distribution, meaning that

\begin{equation}\label{eq1}
p(\theta|\mathcal{D}_{\rm L}^{'}) \approx p(\theta|\mathcal{D}_{\rm H}^{'}),
\end{equation}

which makes the transferability and extrapolation of these ML models possible.

By designing NN architectures ($h \in \mathcal{H}$), one can capture the physics of bifurcation and morphological features from simple geometries with low complexity.
Models trained on these data, characterized by parameter distributions $p(\theta|\mathcal{D}_{\rm L}^{\prime})$, can then be extrapolated to predict brain morphological development with a much more complex geometry.

In FEA, morphological data are meshed into discretized tetrahedral elements.
The representation can be directly translated into graphs, where the nodes correspond to the vertices of the elements, and the edges correspond to the edges of the elements.
Graph neural networks (GNN) are then be constructed to extract key features from the graphs.
We utilize an encoder-decoder architecture to learn the complexity of morphological development (\zfig{Fig. S4}).
The input to the model is a graph representation of the morphology, with node features such as the coordinates and normal directions.
The output is the local curvatures.
The encoder employs three message-passing blocks to transform the input features, comprising 3D positions and normals ($6$-dimensional), into a latent representation with $256$ dimensions.
In the first message-passing block, a multilayer perceptron (MLP) processes the $6$D input using $64$ neurons in both the hidden and output layers.
This is followed by two additional message-passing blocks with $128$ and $256$ neurons, respectively, progressively increasing the expressive capacity of the node features.
After message passing, several fully connected layers further refine the learned representations.
A global feature is produced via a fully connected layer ($256$ to $1$ neuron) followed by global pooling.
Local curvature is predicted using a separate $256$-to-$1$ neuron layer, and a $256$-to-$512$ layer encodes the final latent geometry representation.
The decoder maps this $512$-dimensional latent code to reconstruct the 3D coordinates using an MLP with hidden layers of sizes $512$, $256$, $128$, and $32$, culminating in a 3D output.
Training is guided by a composite loss function that incorporates errors from the reconstructed 3D coordinates, global gyrification index, and local curvatures, with all neural network parameters updated using standard backpropagation.
The adaptive moment estimation (ADAM) optimizer is used to train the neural network parameters, with hyperparameter settings: a learning rate of $0.001$, and first- and second-moment decay rates $\beta_0 = 0.9, \beta_1 = 0.999$.
The algorithm is implemented in PyTorch, and the hyperparameters are set using widely adopted configurations from prior works rather than through extensive tuning~\cite{pfaff2020ICLR}.


\subsection*{Collection of medical data}

We collect experimental data on the human brain to validate our PT approach.
The data of human brain morphologies are rare, especially for the sequences of individual brain morphologies (\zfig{Fig. S2})~\cite{bethlehem2022Nature,ciceri2024NI}.
We use high-resolution MRI data of brain anatomy from open-source brain structural atlases~\cite{ciceri2024NI}, which are then translated to 3D model geometries using a pipeline involving cortical and sub-cortical volume segmentation and cortical surface extraction~\cite{makropoulos2018NI}.

Specifically, we used a triangular surface mesh representation of the human cortical surface, consisting of $32,492$ vertices per hemisphere, derived from a population-averaged atlas covering $21$ to $36$ weeks of gestation (\url{https://doi.gin.g-node.org/10.12751/g-node.qj5hs7/})~\cite{karolis2023atlas}.

\subsection*{Distance metrics in latent spaces}
To assess the generalization ability of a model prior to its deployment, we compute the distance between training samples (sphere data) and testing samples (brain data) in the latent space.
The underlying hypothesis is that a greater distance indicates a larger domain shift, which typically leads to poorer model generalization. Therefore, this distance can be used to estimate model performance \emph{a priori}.

To quantify the distance between the training and testing distributions, we first project both datasets into a low-dimensional latent space using principal component analysis (PCA)~\cite{pearson1901PCA}, which retains the most significant features while reducing noise and redundancy.
Next, we apply kernel density estimation (KDE) to the training data to estimate its probability density function in the latent space~\cite{parzen1962KDE,zong2018ICLR}.
Given a test sample $\tilde{q}$ and a set of $N$ projected training samples $\{\tilde{p}_i\}_{i=1}^N$, the estimated density at $\tilde{q}$ is calculated as

\begin{equation}
\hat{f}(\tilde{q}) = \frac{1}{N}\sum_{i=1}^N K\left(\frac{\|\tilde{q} - \tilde{p}_i\|}{h}\right),
\end{equation}

where $K(\cdot)$ is the Gaussian kernel function defined as

\begin{equation}
K(u) = \frac{1}{2\pi} \exp\left(-\frac{1}{2}u^2\right).
\end{equation}

Here, $h$ is the bandwidth parameter controlling the smoothness of the density estimate.
A lower estimated density $\hat{f}(\tilde{q})$ implies a larger distance from the training distribution and suggests a higher likelihood of prediction error.

This KDE-based distance measure provides a quantitative approach to evaluate the proximity of testing data to the training distribution.
It enables the estimation of prediction uncertainty and can serve as an early warning indicator for potential model failure due to domain shift.

\clearpage
\newpage

\section*{Declaration of Competing Interest}

\noindent The authors declare that they have no competing financial interests.

\section*{Credit authorship contribution statement}

\noindent Z.X. conceived the concept.
Y.Z. developed the methods and performed the research under the supervision of Z.X. and F.X.
All authors analysed the results and wrote the manuscript.




\section*{Acknowledgements}
 
\noindent This work was supported by the National Natural Science Foundation of China through grants 12425201, 52090032, 12125206, 124B2035, 12425204, 12122204 and 12372096, Shanghai Pilot Program for Basic Research-Fudan University (grant no. 21TQ140010021TQ010), and Shanghai Municipal Education Commission (grant no. 24KXZNA14).
The computation was performed on the Explorer 1000 cluster system of the Tsinghua National Laboratory for Information Science and Technology.


\clearpage
\newpage

\bibliography{main_text}
\clearpage
\newpage

\section*{Figures and Figure Captions}

\zfig{\bf Figure 1.} Brain development complexity and challenges arising from experimental low fidelity and data scarcity.

\zfig{\bf Figure 2.} Physics-transfer (PT) approach for brain morphological development.

\zfig{\bf Figure 3.} Descriptive and predictive performance of PT models. 

\zfig{\bf Figure 4.} Reduced-dimensional models and neural network analysis.

\zfig{\bf Figure 5.} Medical applications and the perspective demonstration.

\clearpage
\newpage

\begin{figure}[H]
\centering
\includegraphics[width=\linewidth] {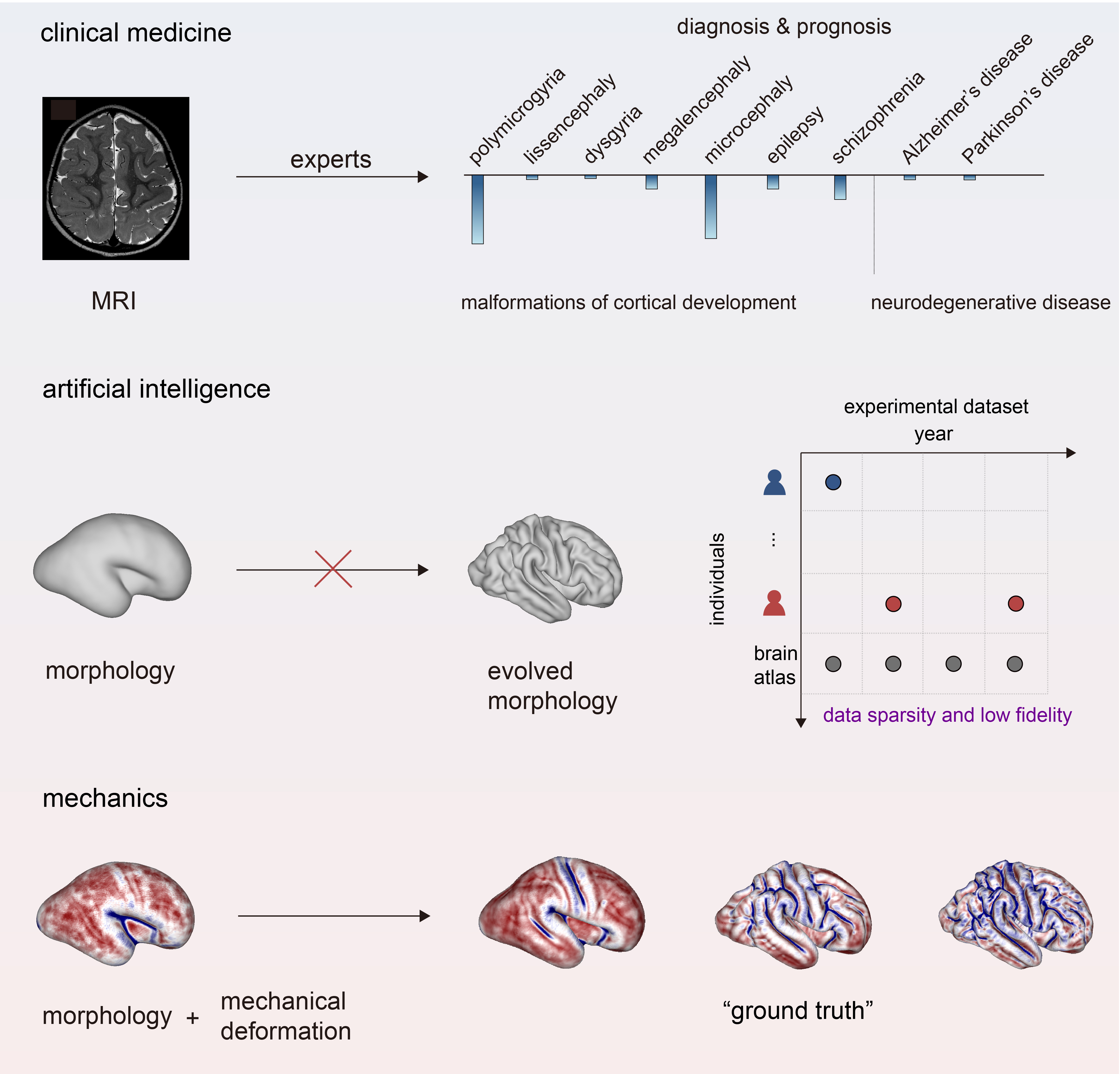} 
\caption*{{\bf Figure 1. Complexity in brain development and challenges from experimental low fidelity and data scarcity.}
Current diagnostic practices primarily rely on expert interpretation of magnetic resonance imaging (MRI) scans, which provide geometric information based on voxel intensity but lack biomechanical fidelity.
Moreover, the majority of MRI data in both clinical and experimental contexts are cross-sectional, limiting the ability to investigate temporal morphological changes.
This scarcity of longitudinal data poses a significant barrier to the deployment of AI-driven methods for modeling brain development or disease progression.
By leveraging continuum mechanics, it is possible to reconstruct high-fidelity stress fields from observed geometries, offering a pathway to more comprehensive insights into brain morphogenesis.
}
\label {fig1}
\end{figure}

\clearpage
\newpage

\begin{figure}[H]
\centering
\includegraphics[width=\linewidth] {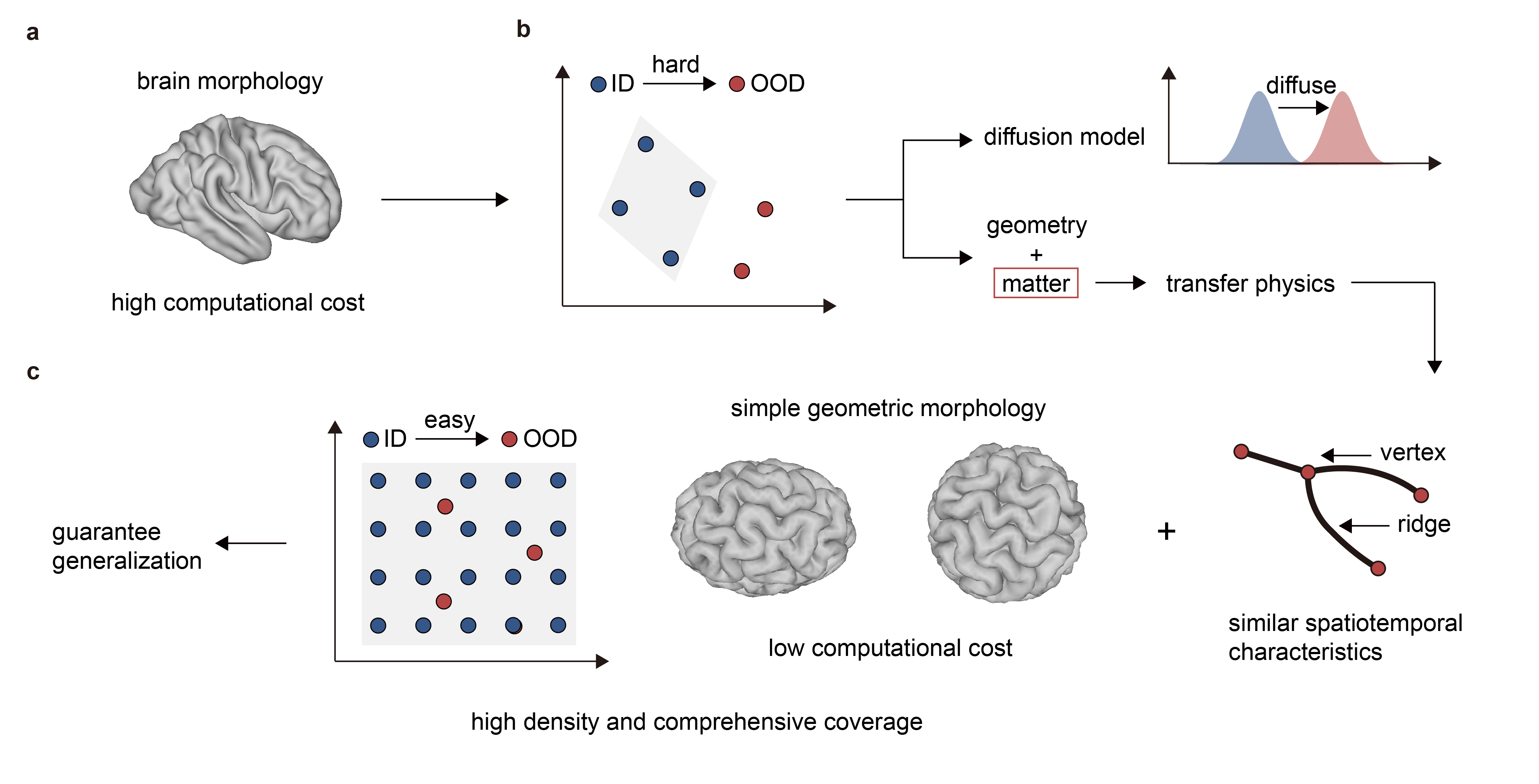} 
\caption*{{\bf Figure 2. Physics-transfer (PT) approach to predicting brain morphological development.}
{\bf (a)} The inherent geometric complexity of brain anatomy poses a significant challenge to constructing a comprehensive digital library from limited MRI datasets, resulting in data sparsity.
ID: in distribution.
OOD: out of distribution.
{\bf (b)} While generative AI holds promise, purely data-driven approaches based on statistical learning (SL) from morphological data remain constrained by the low fidelity and scarcity.
{\bf (c)} To address this, we decouple physical and geometric complexity by first modeling morphological development using simplified geometries.
This enables the creation of a dense, well-sampled digital library, from which the learned physics of nonlinear elasticity can be transferred to anatomically complex brain data.
Sampling from the morphological space of simplified geometries proves significantly more efficient than direct sampling from brain tissue models.
These simplified domains provide high coverage and density, and their spatiotemporal similarity to brain morphological evolution supports robust generalization of the underlying physics.
}
\label {fig2}
\end{figure}

\clearpage
\newpage

\begin{figure}[H]
\centering
\includegraphics[width=\linewidth] {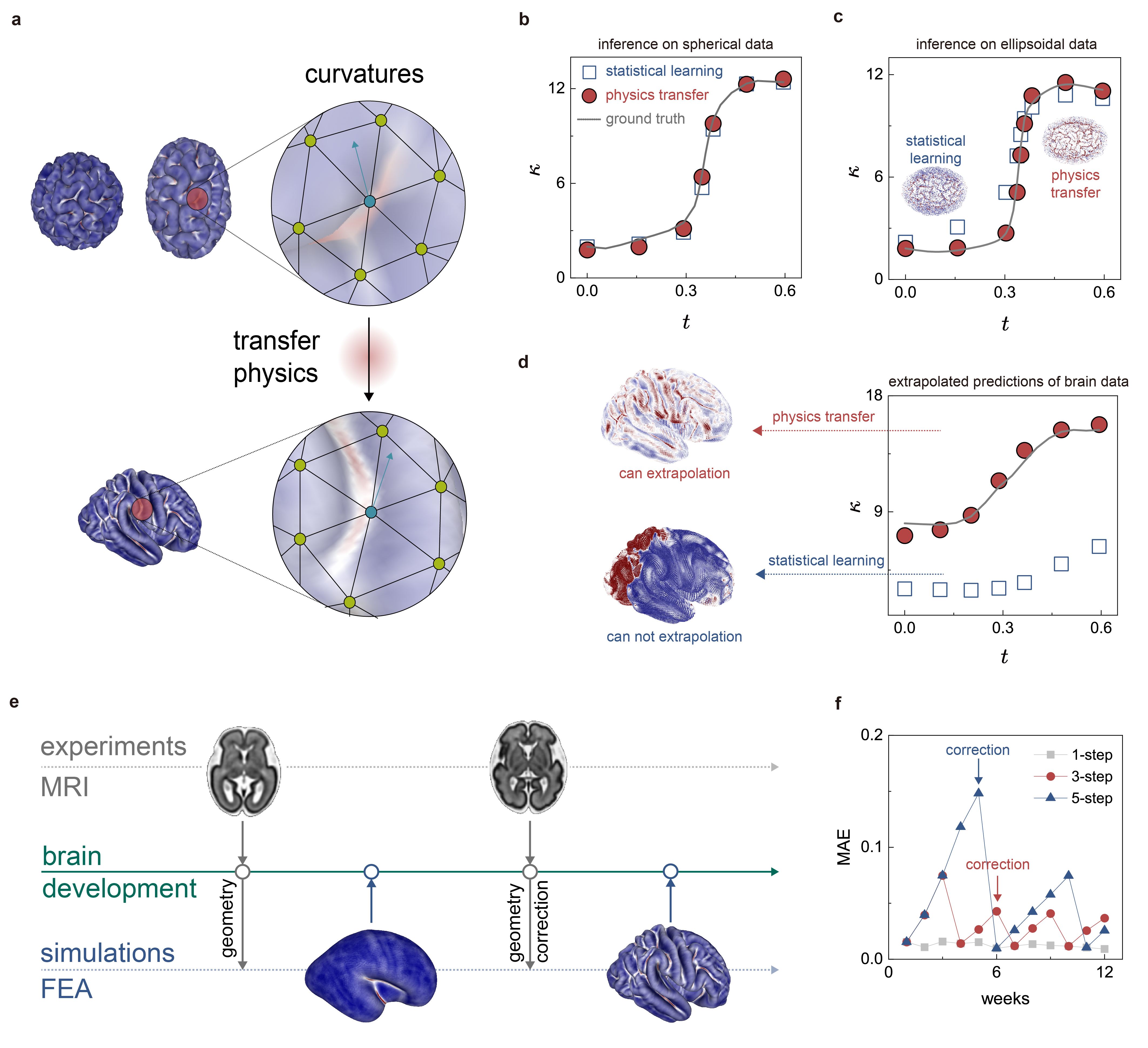} 
\caption*{{\bf Figure 3. Descriptive and predictive performance of PT models.}
{\bf (a)} PT models applied to descriptive curvature maps. 
{\bf (b)} Interpolative predictions for spherical data.
{\bf (c, d)} Extrapolative predictions for ellipsoidal data {\bf (c)} and the development of brain morphologies {\bf (d)}.
The insets are colored by the local curvature metrics, $\lvert H \lvert + \lvert K \lvert$, where $H$ and $K$ are the mean and Gaussian curvatures, respectively.
{\bf (e)} PT models applied to predictive morphological development, enabling a digital twin approach for medical applications.
{\bf (f)} Predictive performance of PT models in brain morphological development.
}
\label {fig3}
\end{figure}

\clearpage
\newpage

\begin{figure}[H]
\centering
\includegraphics[width=\linewidth] {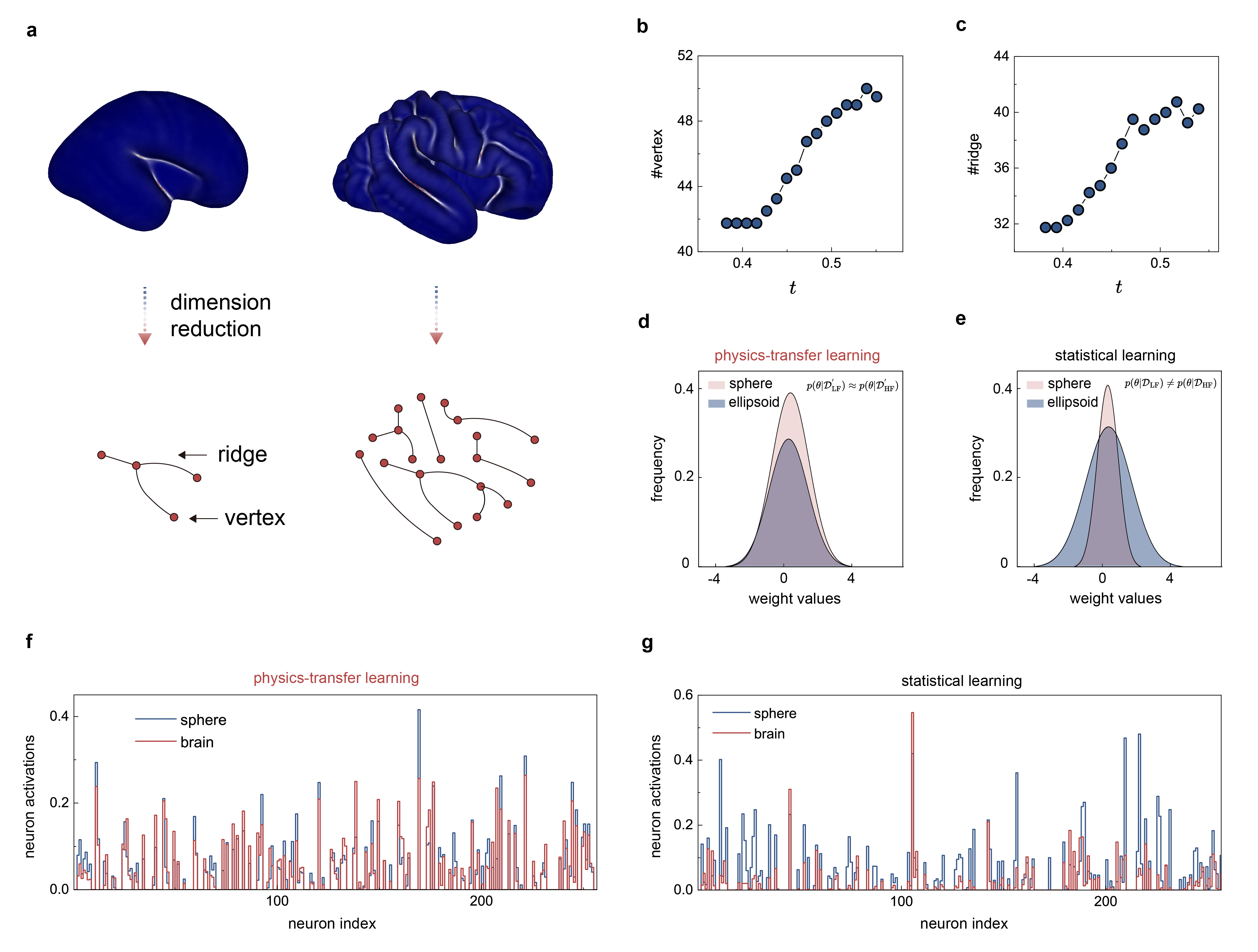} 
\caption*{{\bf Figure 4. Reduced-dimensional models and neural network analysis.}
{\bf (a)} The vertex-ridge network emerging from the machine learning (ML) model.
{\bf (b, c)} Dimension-reduction metrics used to track the evolutionary dynamics in brain morphological development, which include the number of the vertices {\bf (b)} and the number of ridges {\bf (c)}.
{\bf (d, e)} Weights parameters distribution of ML models trained on the spherical and ellipsoidal data for PT learning {\bf (d)} and SL with only the morphological data used in the learning process {\bf (e)}.
{\bf (f, g)} Neuron activations of the ML models trained on spherical data, when inference on both spherical and brain data for PT learning {\bf (f)} and SL {\bf (g)}.
}
\label {fig4}
\end{figure}

\clearpage
\newpage

\begin{figure}[H]
\centering
\includegraphics[width=\linewidth] {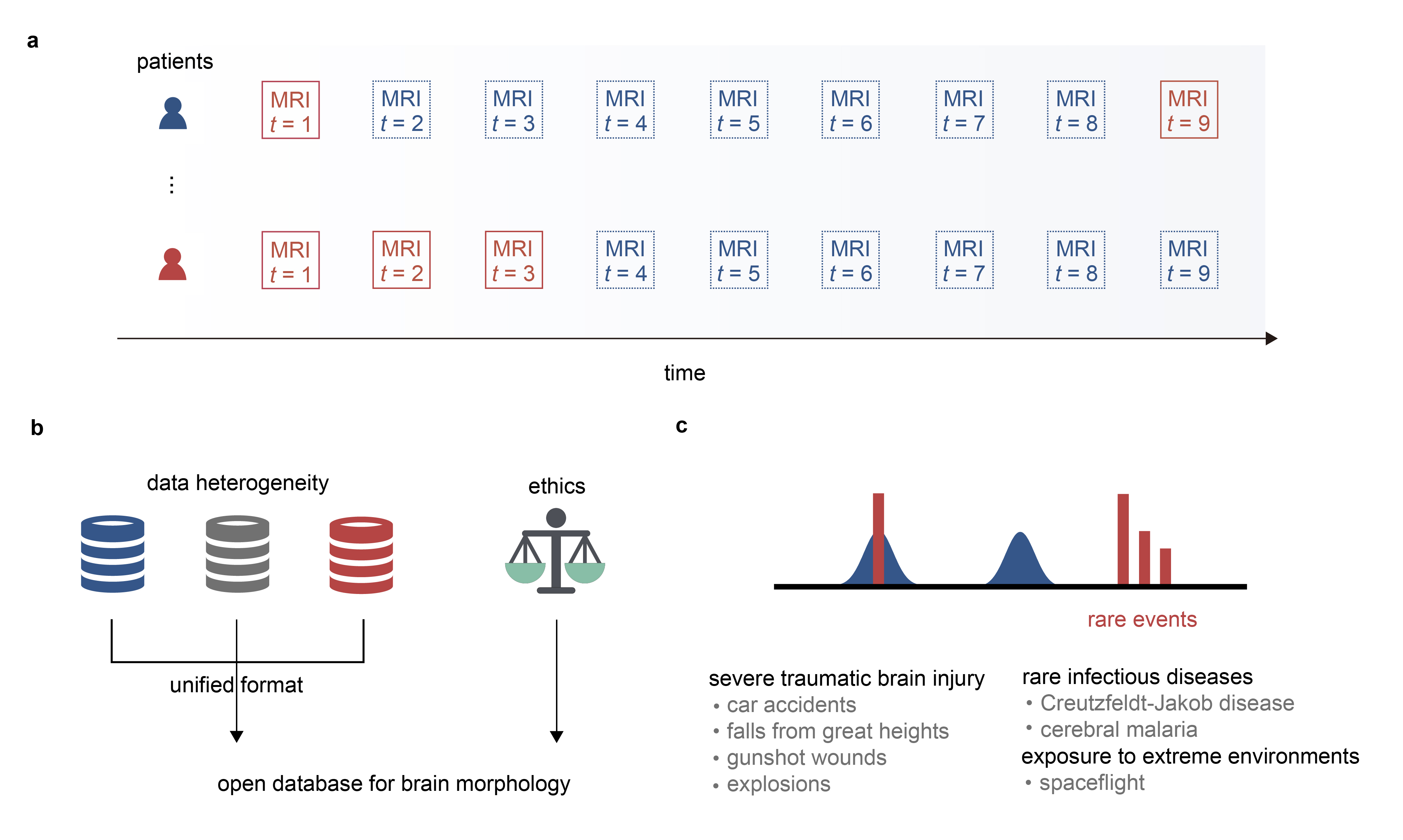} 
\caption*{{\bf Figure 5. Medical applications and the perspective demonstration.}
{\bf (a)} Longitudinal MRI data enable personalized prediction of brain morphological development using PT models.
{\bf (b)} Challenges in building an integrated open database for brain development. 
{\bf (c)} Effects of rare events on brain morphogenesis and disease.
}
\label {fig5}
\end{figure}

\clearpage
\newpage

\section*{Supplementary Material for Predicting Brain Morphogenesis via Physics-Transfer Learning by Zhao et al.}

This {\bf Supplementary Material} includes {\bf Supplementary Notes S1-S2}, {\bf Supplementary Figures S1-S7}, and {\bf Figure Captions}.

\zfig{\bf Note S1.} Reduced-dimensional analysis.

\zfig{\bf Note S2.} Information bottleneck (IB) analysis.

\zfig{\bf Figure S1.} Accuracy-performance dilemma in multiscale modeling of brain morphological development.

\zfig{\bf Figure S2.} Experimental magnetic resonance imaging (MRI) datasets and finite element analysis (FEA) digital libraries of brain morphological development.

\zfig{\bf Figure S3.} Physics-transfer (PT) framework that learns physics across models of varying geometric complexities. 

\zfig{\bf Figure S4.} The neural network architecture.

\zfig{\bf Figure S5.} Schematic diagram of multi-step autoregressive prediction.

\zfig{\bf Figure S6.} Distribution differences of latent features for assessing model generalization.

\zfig{\bf Figure S7.} The information bottleneck theory and PT model distinguish the effects of curvature at different scales.

\clearpage
\newpage

\section*{Supplementary Note S1. Reduced-dimensional analysis}

We extract the reduced-dimensional representations from morphological data by identifying key feature points (vertices) and connecting paths (ridges) across temporal frames.
The raw morphological data is first filtered based on domain-specific criteria (e.g., intensity and thresholding).
Density-based spatial clustering of applications with noise (DBSCAN) is implemented to remove noise and isolate ridge-relevant points, which represent the skeletal structure of the morphology~\cite{ester1996KDD}.
A $k$-nearest neighbor graph is constructed over the cleaned ridge points.
Nodes with a degree of $1$ (endpoints) or $3$ and above (junctions) are initially considered as vertex candidates.
To account for spatial uncertainty, nearby candidate points are merged using a distance tolerance, resulting in a refined set of vertices representing biologically meaningful endpoints and bifurcations in the structure.
Ridges are defined as spatially valid connections between vertices.
For each time step, the final set of vertices and ridges forms a simplified graph representation of the morphology.
The number of vertices and ridges is recorded to quantify the evolving complexity of the network over time.

To further assess the structure of NN-based models, we incorporate tools from complex network theory~\cite{la2024PR}.
These tools enable reduced-dimensional analysis of deep neural networks (DNNs), offering insight into their learned representations~\cite{la2024PR}.
A DNN can be represented as a directed graph $G = \langle N, E \rangle$, where nodes $n_{i}^{[\ell]} \in N$ correspond to neurons in the $\ell$-th hidden layer, and edges $(e_{n_{i}^{[\ell]}, n_{j}^{[\ell+1]}} \in E)$ represent weighted connections with real-valued weights $\omega_{i,j}^{[\ell]}$.
A standard set of metrics involves computing the mean and variance of the weights in each layer of the network. For the $\ell$-th layer of a DNN, these are defined as

\begin{equation}
\mu^{[\ell]}=\frac{1}{N^{[\ell]}N^{[\ell+1]}}\sum_{i=1}^{N^{[\ell]}}\sum_{j=1}^{N^{[\ell+1]}}(\omega_{i,j}^{[\ell]}+\beta_{i}^{[\ell]}),
\end{equation}

\begin{equation}
\delta^{[\ell]}=\frac{1}{N^{[\ell]}N^{[\ell+1]}}\sum_{i=1}^{N^{[\ell]}}\sum_{j=1}^{N^{[\ell+1]}}\left((\omega_{i,j}^{[\ell]}+\beta_{i}^{[\ell]})-\mu^{[\ell]}\right)^2.
\end{equation}
Analyzing these metrics with varying geometric complexity, one can assess the model's generalization capability.

Resolving neuron activation patterns in the latent space provides a means to assess the model's uncertainty in generalizing to out-of-distribution (OOD) samples.
The activation value of the $k$-th neuron in the $\ell$-th layer is given by:

\begin{equation}
a_{k}^{[\ell]} = f^{[\ell]}\left(\sum_{i=1}^{N^{[\ell]}}z_{i}^{[\ell-1]}\omega_{i,k}^{[\ell]}+\beta_{k}^{[\ell]}\right),
\end{equation}
\begin{equation}
z^{[\ell-1]}=\mathbf{f}(x,\Omega^{[\cdot\ell]},\beta^{[\cdot\ell]}),\quad x\sim\mathcal{X}.
\end{equation}

These equations indicate that the activation $a_{k}^{[\ell]}$ depends on the weighted sum of activations $z_{i}^{[\ell-1]}$ from the previous layer, modulated by weights $\omega_{i,k}^{[\ell]}$ and bias $\beta_{k}^{[\ell]}$, followed by a nonlinear transformation via the activation function $f^{[\ell]}$.
For samples with high uncertainty, neurons in the latent space are expected to exhibit abnormally high node strengths, indicating a significant deviation from the activation patterns learned during training.
PT models display consistent activation patterns between spherical and brain data compared to statistical learning (SL) models, indicating superior generalization and reduced uncertainty.

\clearpage
\newpage

\section*{Supplementary Note S2. Information bottleneck (IB) analysis}

To further disentangle the influence of configuration versus deformation curvature, we construct two datasets: one from morphogenetic processes on flat plates (lacking configuration curvature) and another from spherical geometries (with inherent configuration curvature).
Monitoring information compression reveals greater compression on the spherical dataset ($\sim 34\%$) compared to the flat plate dataset ($\sim 31\%$) (\zfig{Fig. S7d}), suggesting that deformation curvature contributes more critically to the spatiotemporal complexity of morphogenesis by providing richer, more informative structural features for the model to learn.

\clearpage
\newpage
\renewcommand\thefigure{S\arabic{figure}}
\addtocounter{figure}{-5}

\begin{figure}[htp]
\centering
\includegraphics[width=\linewidth] {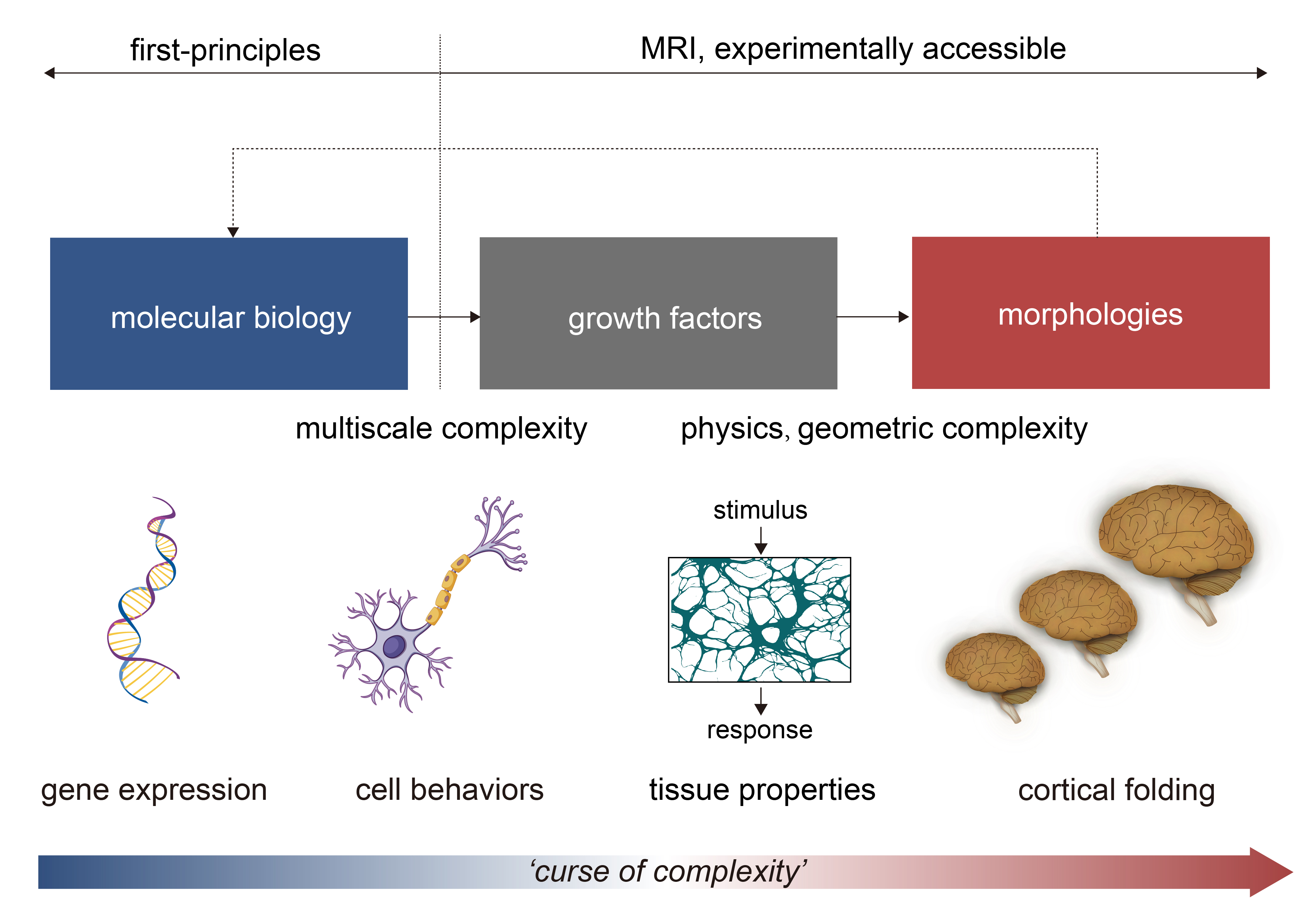} 
\caption*{{\bf Figure S1. Accuracy-performance dilemma in multiscale modeling of brain morphological development.}
} 
\end{figure}

\clearpage
\newpage

\begin{figure}[htp]%
\centering
\includegraphics[width=\linewidth] {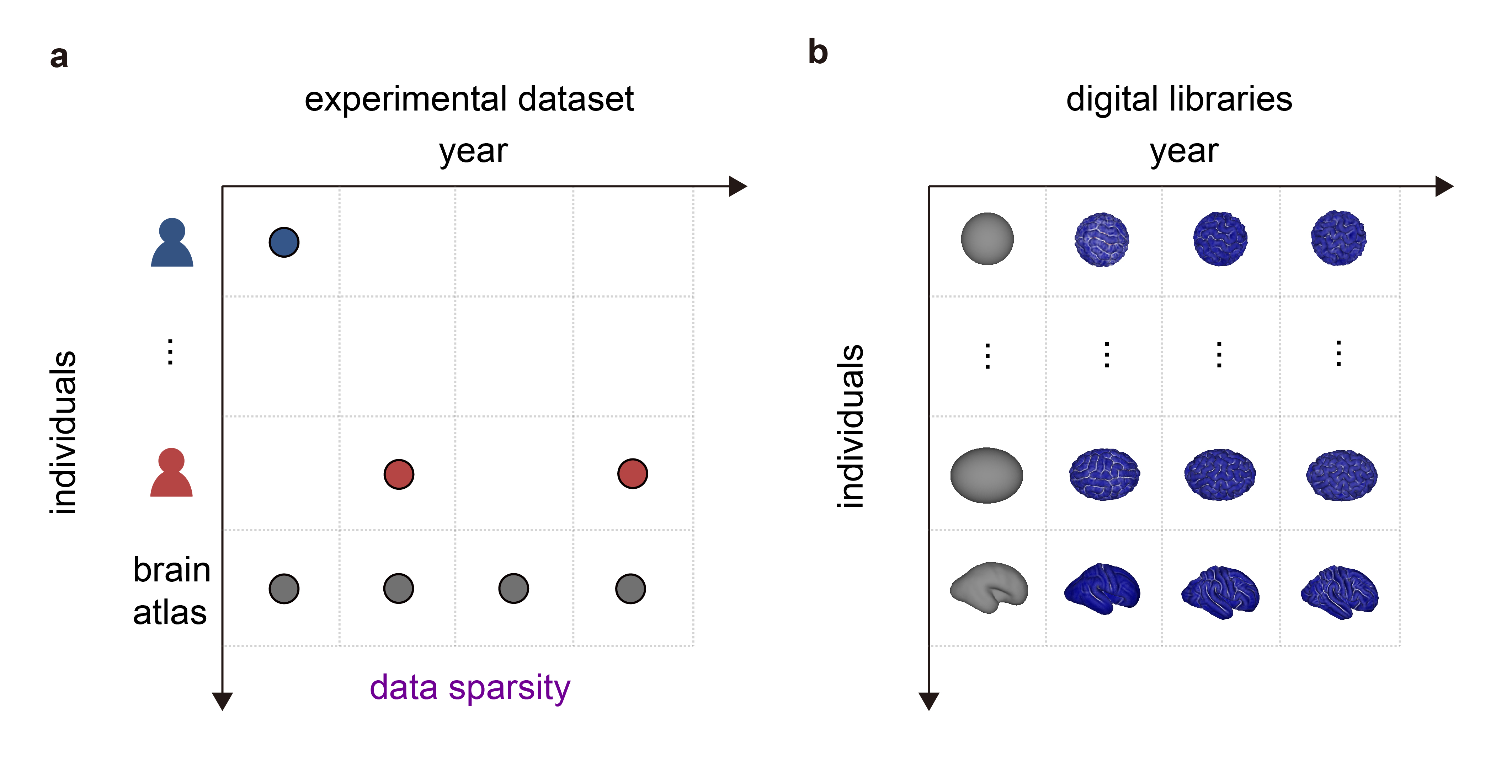} 
\caption*{{\bf Figure S2. Experimental magnetic resonance imaging (MRI) datasets and finite element analysis (FEA) digital libraries of brain morphological development.}
{\bf (a)} Experimental datasets collected from the literature~\cite{ciceri2024NI}. 
{\bf (b)} Digital libraries constructed from our FEA.
}
\end{figure}

\clearpage
\newpage

\begin{figure}[htp]
\centering
\includegraphics[width=\linewidth] {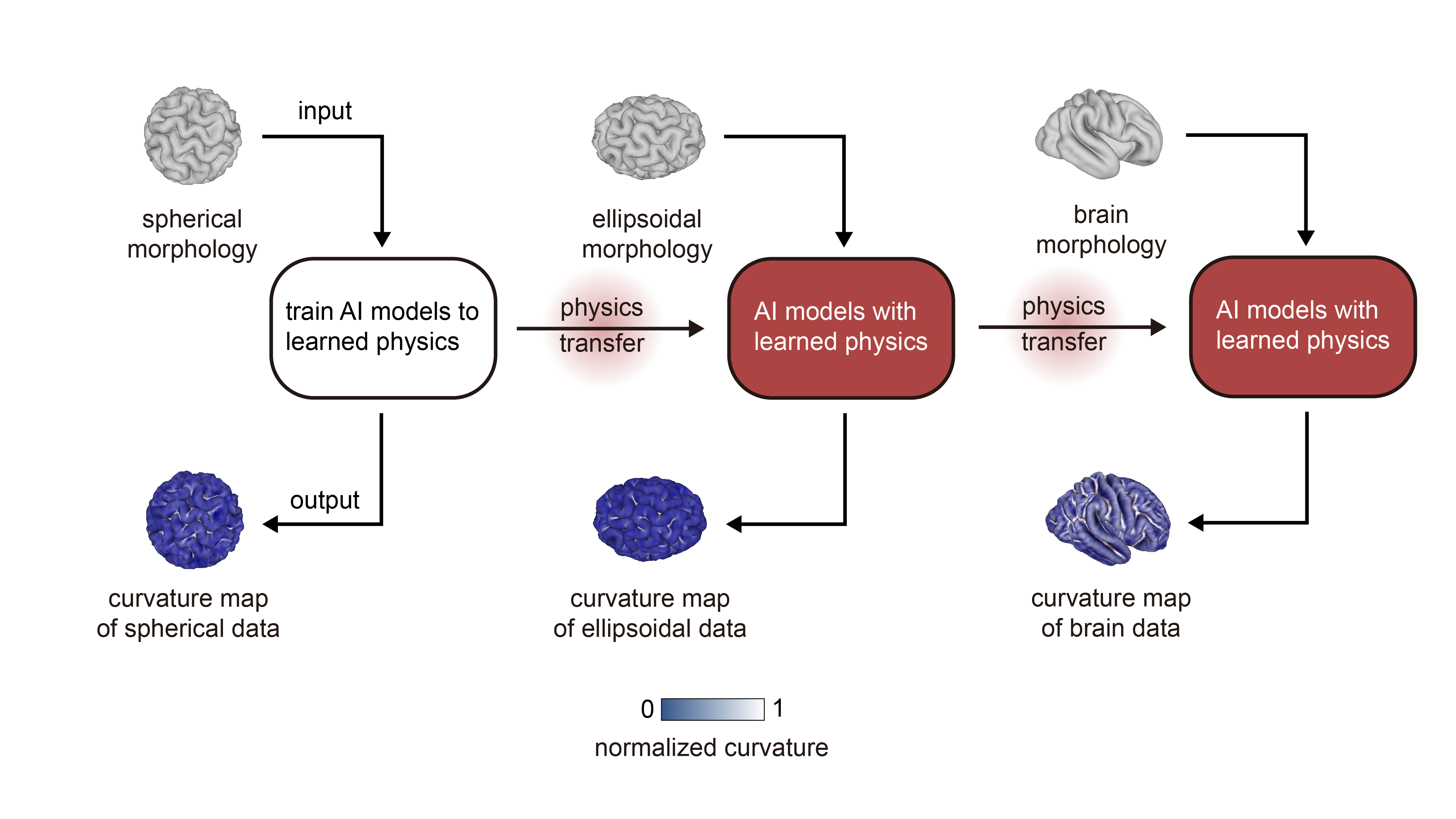} 
\caption*{{\bf Figure S3. Physics-transfer (PT) framework that learns the physics of developmental nonlinear deformation across models of varying geometric complexities.}
}
\end{figure}

\clearpage
\newpage

\begin{figure}[htp]%
\centering
\includegraphics[width=\linewidth] {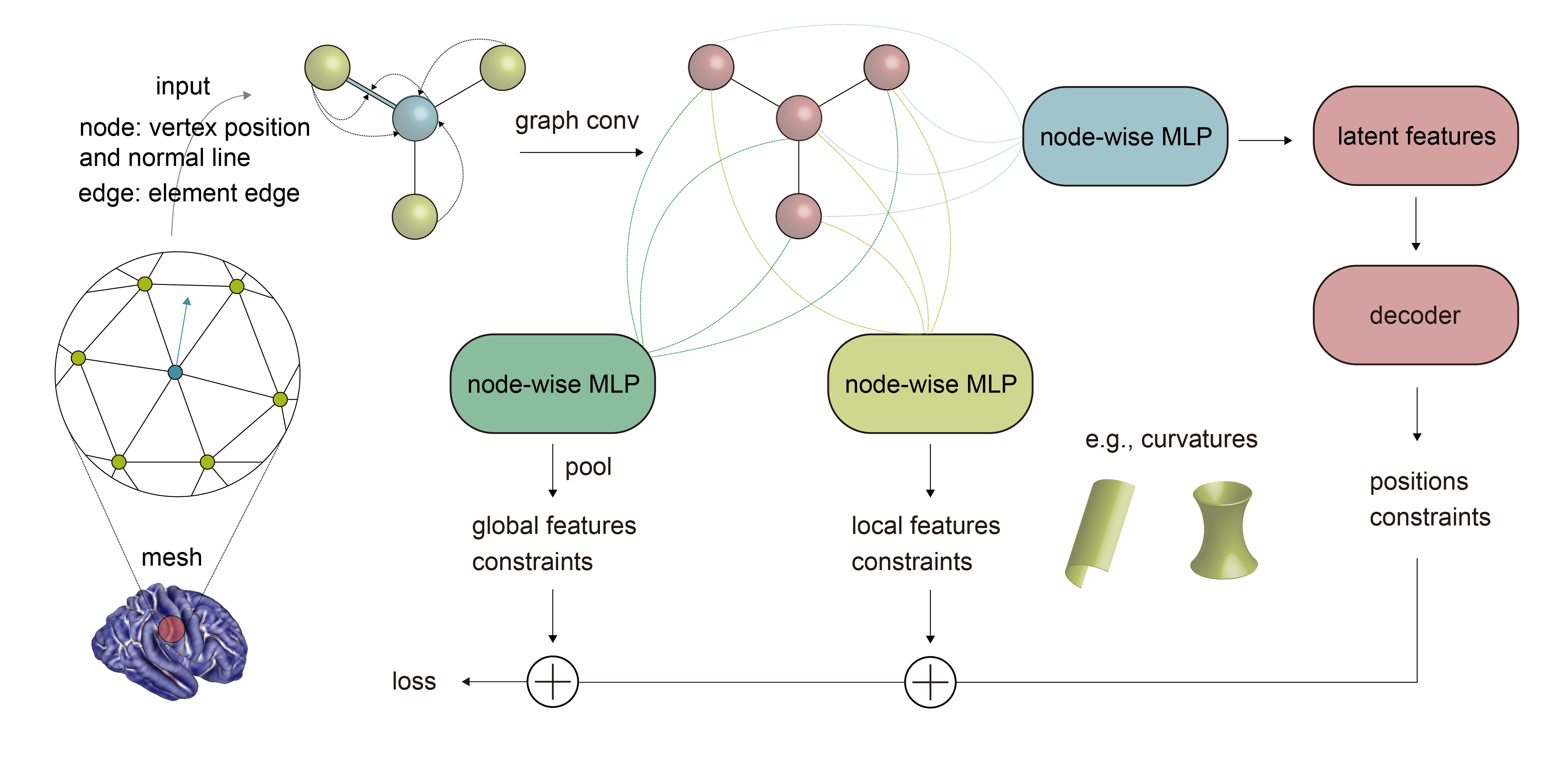} 
\caption*{{\bf Figure S4. The neural network architecture.}
An encoder-decoder architecture is employed to capture the complexity of morphological development.
The model takes as input a graph representation of the morphology, where node features include coordinates and normal directions.
The output is local curvatures.
Additionally, the 3D coordinates of the morphologies and global features, such as the gyrification index, are incorporated into the loss function to constrain the model.} 
\end{figure}

\clearpage
\newpage

\begin{figure}[htp]%
\centering
\includegraphics[width=\linewidth] {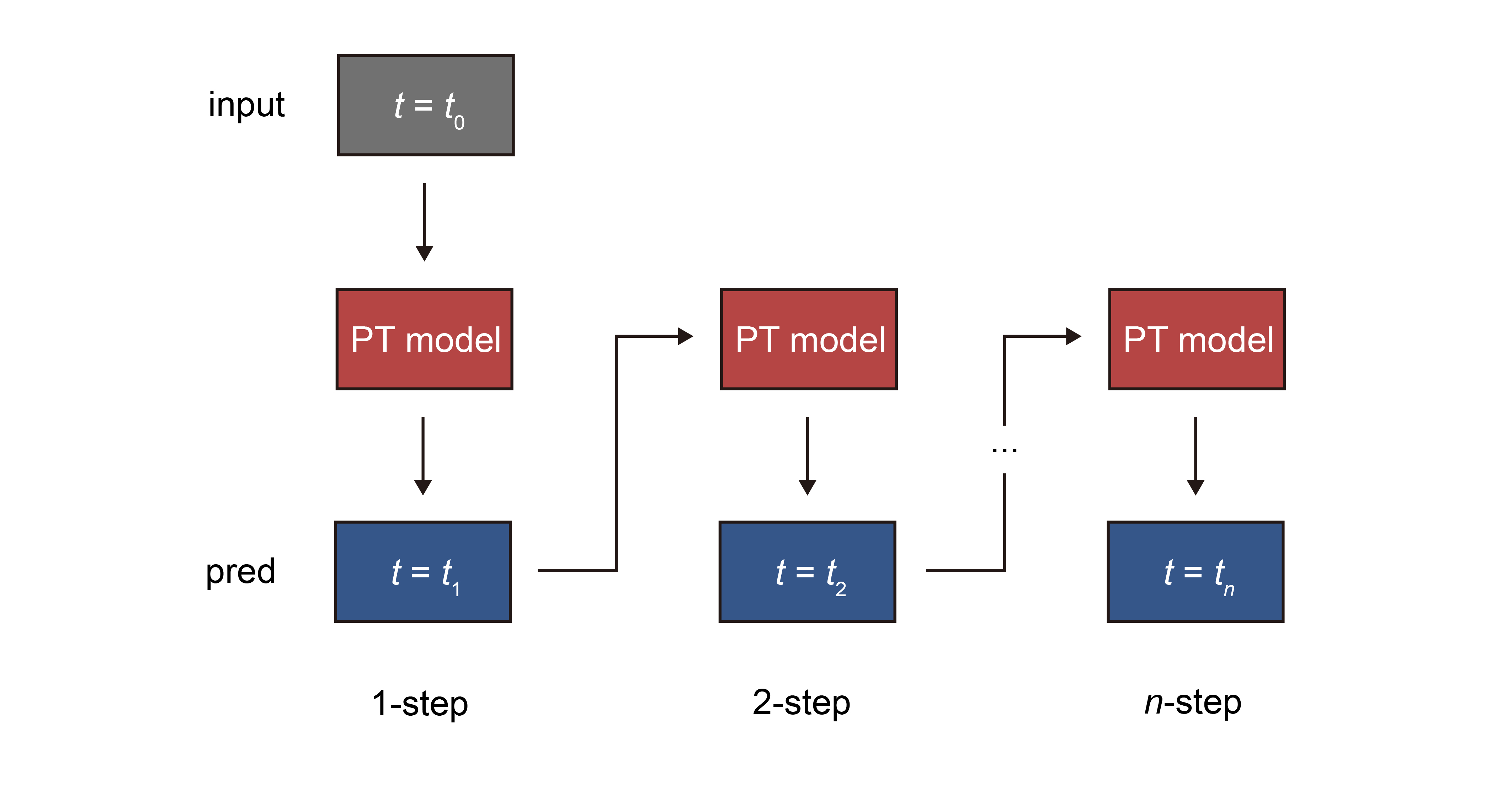} 
\caption*{{\bf Figure S5. Schematic diagram of multi-step autoregressive prediction.}
}
\end{figure}

\clearpage
\newpage

\begin{figure}[htp]%
\centering
\includegraphics[width=\linewidth] {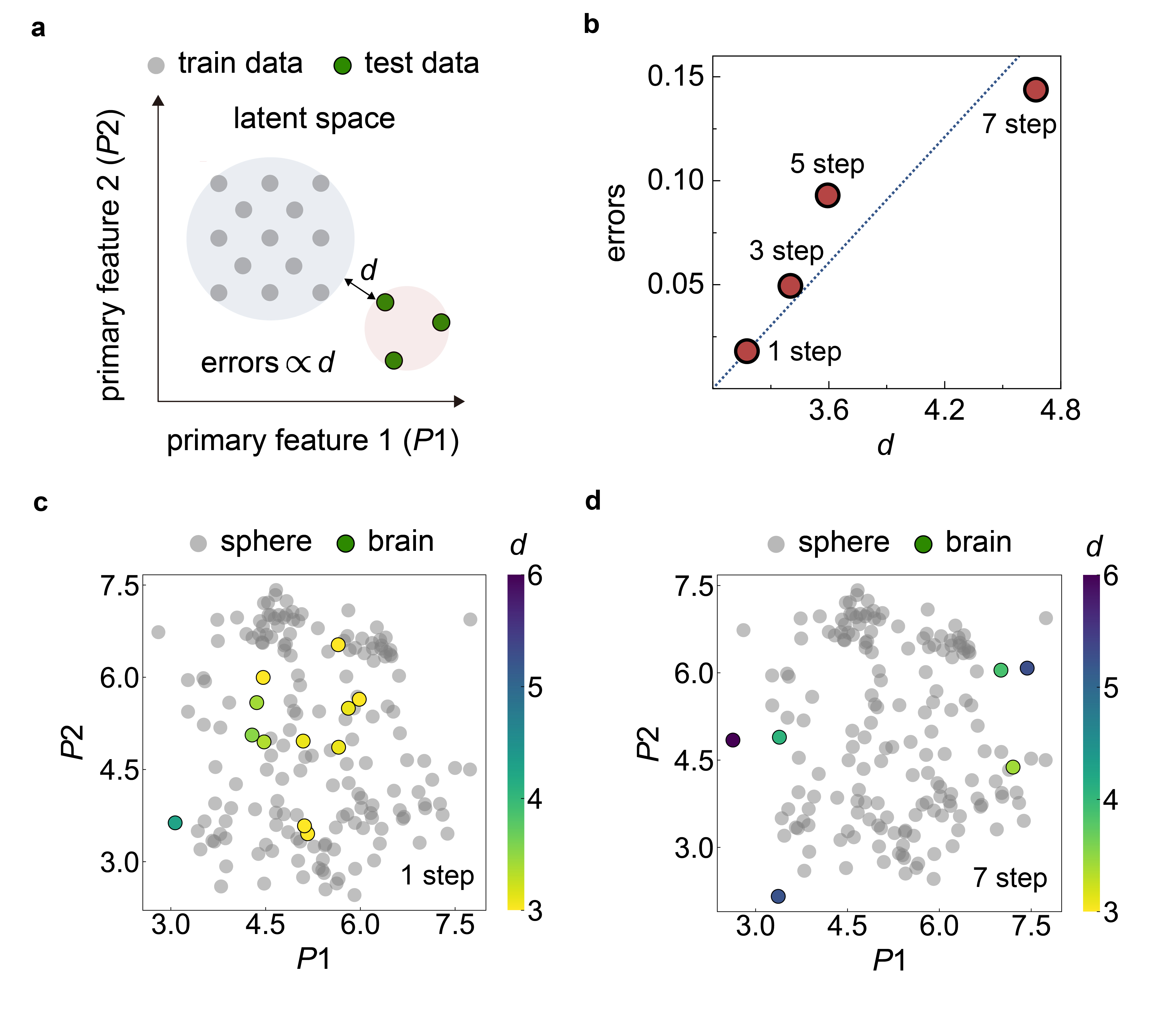} 
\caption*{{\bf Figure S6. Distribution differences of latent features for assessing model generalization.}
{\bf (a)} Quantifying the relationship between latent space features and generalizability for prior estimation of model performance. 
{\bf (b)} Prediction error increases with the distance between training spherical data and testing brain data, indicating reduced generalizability with greater dissimilarity.
{\bf (c,d)} In the latent space, the $1$-step brain input is closest to the sphere dataset and yields the lowest prediction error {\bf (c)}, whereas the $7$-step input is farthest and has the highest error, illustrating reduced generalization with increasing input complexity {\bf (d)}.
}
\end{figure}

\clearpage
\newpage

\begin{figure}[htp]%
\centering
\includegraphics[width=\linewidth] {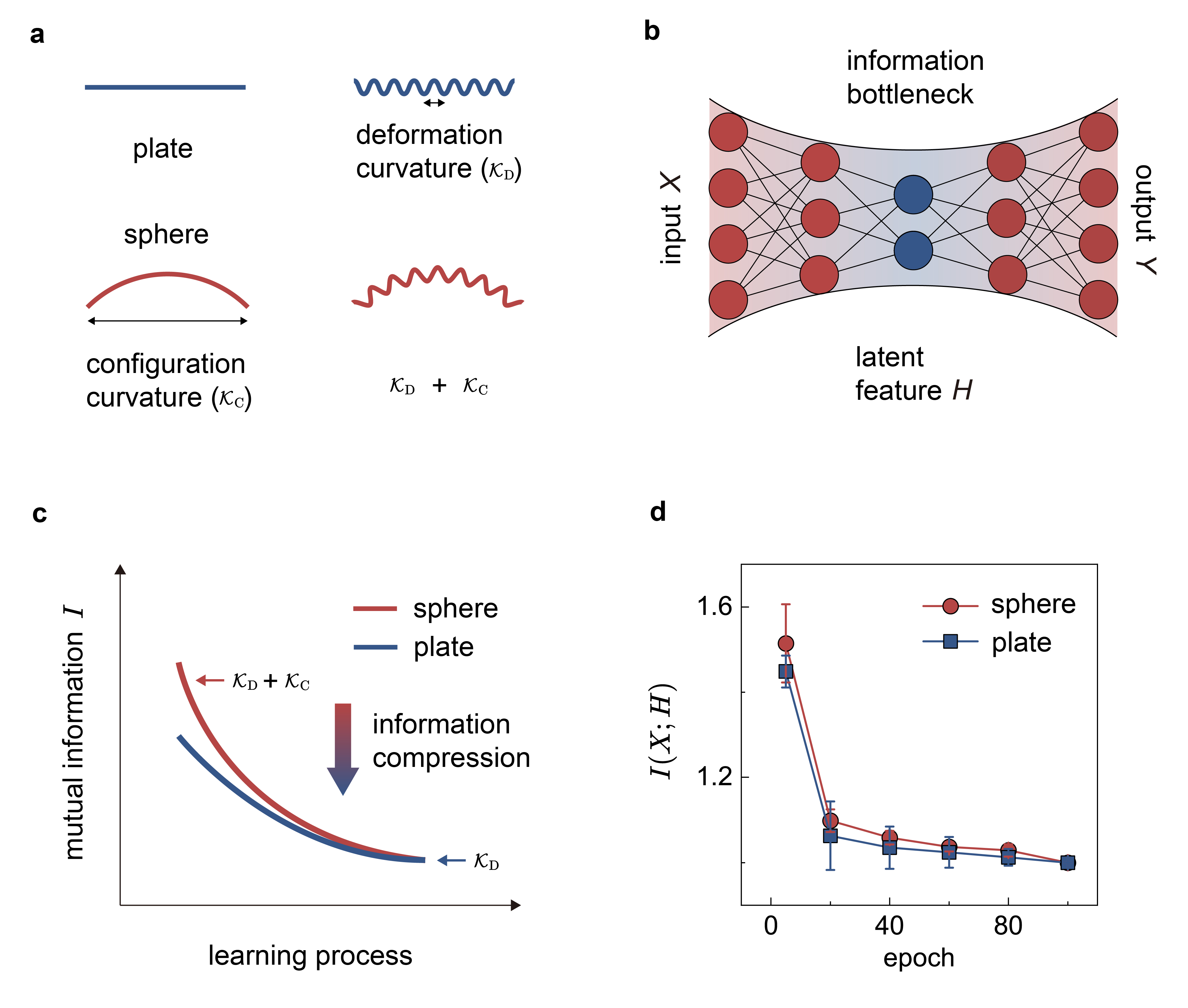} 
\caption*{{\bf Figure S7. The information bottleneck theory and PT model distinguish the effects of curvature at different scales.}
{\bf (a)} Curvatures at different scales during morphogenesis.
{\bf (b)} Deep neural network and information bottleneck framework.
{\bf (c)} Convergence of mutual information during training reflects the extraction of latent features that capture morphogenetic principles.
{\bf (d)} Information compression during training for the spherical dataset and the flat plate dataset.
}
\end{figure}

\end{document}